\documentclass{article}
\usepackage{amssymb}
\begin{document}
\title{Variational symmetries as the existence of ignorable coordinates}

\author{G.F.\ Torres del Castillo \\ Departamento de F\'isica Matem\'atica, Instituto de Ciencias \\
Universidad Aut\'onoma de Puebla, 72570 Puebla, Pue., M\'exico \\[2ex]
I.\ Rubalcava Garc\'ia \\ Facultad de Ciencias F\'isico Matem\'aticas \\ Universidad Aut\'onoma de Puebla, 72570 Puebla, Pue., M\'exico}

\maketitle

\begin{abstract}
It is shown that given a Lagrangian for a system with a finite number of degrees of freedom, the existence of a variational symmetry is equivalent to the existence of coordinates in the extended configuration space such that one of the coordinates is ignorable.
\end{abstract}

\section{Introduction}
In the application of the Lagrangian formalism in classical mechanics, the existence of an ignorable coordinate in the Lagrangian of the system under study leads to a conserved quantity (the momentum canonically conjugated to the ignorable coordinate). This connection between ignorable coordinates and conserved momenta follows in a straightforward manner from the Euler--Lagrange equations, and is highlighted in almost every textbook on Lagrangian mechanics. However, the existence of ignorable coordinates in a Lagrangian depends on the appropriate choice of the coordinates, for which, usually, one may have no clues.

One of the reasons why the conserved quantities are important is that the solution of the equations of motion is equivalent to the knowledge of an appropriate number of functionally independent conserved quantities.

There is a way of finding some of the conserved quantities admitted by (the equations of motion derived from) a given Lagrangian, which is usually obtained making use of the concept of a one-parameter group of variational symmetries of a Lagrangian. Here we only summarize the basic result: If $L(q_{i}, \dot{q}_{i}, t)$ is a Lagrangian for a system with $n$ degrees of freedom and the $n + 1$ functions $\eta_{i}(q_{j}, t)$, $\xi(q_{j}, t)$ satisfy the partial differential equation
\begin{equation}
\sum_{i = 1}^{n} \left[ \frac{\partial L}{\partial q_{i}} \eta_{i} + \frac{\partial L}{\partial \dot{q}_{i}} \left( \frac{{{\rm d}} \eta_{i}}{{{\rm d}} t} - \dot{q}_{i} \frac{{{\rm d}} \xi}{{{\rm d}} t} \right) \right] + \frac{\partial L}{\partial t} \xi + L \, \frac{{{\rm d}} \xi}{{{\rm d}} t} = 0,  \label{invinfs}
\end{equation}
where we have made use of the abbreviation ${\rm d} f/{\rm d} t := \partial f/\partial t + \sum_{i =1}^{n} \dot{q}_{i} (\partial f/\partial q_{i})$, for any function $f(q_{i}, t)$, then
\begin{equation}
\varphi(q_{i}, \dot{q}_{i}, t) = \sum_{i = 1}^{n} \frac{\partial L}{\partial \dot{q}_{i}} \eta_{i} + \xi \left( L - \sum_{i = 1}^{n} \frac{\partial L}{\partial \dot{q_{i}}} \dot{q}_{i} \right) \label{conss}
\end{equation}
is a constant of motion (see, e.g., \cite{HS,PO, PH,GFTC-AndradeMiron2013} and the references cited therein). (In fact, a straightforward computation using the Euler--Lagrange equations, shows that the time derivative of the expression in the right-hand side of equation (\ref{conss}) is equal to zero if and only if equation (\ref{invinfs}) holds.)

The aim of this article is to show that the existence of a nontrivial solution of equation (\ref{invinfs}) is equivalent to the existence of a coordinate system in the extended configuration space such that one of the coordinates is ignorable. In other words, we want to show that $L$ possesses a variational symmetry if and only if there exists a coordinate transformation $q'_{i} = q'_{i}(q_{j}, t)$, $t' = t'(q_{j}, t)$, such that $q'_{1}$ is ignorable. In Section 2 we give a proof of this equivalence and in Section 3 we consider a more general class of symmetries. In Section 4 we present some explicit examples, where we start from a variational symmetry and we find new coordinates such that one of them is ignorable.

\section{Ignorable coordinates and variational symmetries}
Given a Lagrangian, $L(q_{i}, \dot{q}_{i}, t)$, the coordinate $q_{1}$, say, is ignorable if $\partial L/\partial q_{1} = 0$. Then, by virtue of the Euler--Lagrange equations,
\begin{equation}
\frac{{\rm d}}{{\rm d} t} \frac{\partial L}{\partial \dot{q}_{i}} - \frac{\partial L}{\partial q_{i}} = 0, \label{lageq}
\end{equation}
it follows that the momentum conjugate to $q_{1}$, $p_{1} := \partial L/\partial \dot{q}_{1}$, is conserved, that is
\[
\frac{{\rm d} p_{1}}{{\rm d} t} = 0.
\]
We want to show that equation (\ref{invinfs}) is equivalent to the existence of a coordinate system $(q'_{i}, t')$ in the extended configuration space such that
\begin{equation}
\frac{\partial L'}{\partial q'_{1}} = 0. \label{Ign}
\end{equation}
We have to take into account that if $t$ is replaced by some function $t' = t'(q_{i}, t)$, then the Lagrangian $L$ has to be replaced by
\begin{equation}
L(q_{i}, \dot{q}_{i}, t) \frac{{\rm d} t}{{\rm d} t'} =: L'(q_{i}', \dot{q}_{i}', t'), \label{trla}
\end{equation}
where $\dot{q}'_{i} := {\rm d} q'_{i}/{\rm d} t'$. (This relation can be derived from the equality
\[
\int_{t_{0}}^{t_{1}} L(q_{i}, \dot{q}_{i}, t) {\rm d} t = \int_{t'_{0}}^{t'_{1}} L'(q'_{i}, \dot{q}'_{i}, t') {\rm d} t',
\]
taking into account the fact that the Euler--Lagrange equations determine the stationary points of the functional
\[
\int_{t_{0}}^{t_{1}} L(q_{i}, \dot{q}_{i}, t) {\rm d} t,
\]
with fixed endpoints $((q_{i})_{0}, t_{0})$, $((q_{i})_{1}, t_{1})$ in the extended phase space.)

We start from equation (\ref{Ign}), that is, assuming that
\begin{equation}
\frac{\partial}{\partial q'_{1}} \left( L \frac{{\rm d} t}{{\rm d} t'} \right) = 0 \label{StartingPoint-ConditionSymmetry}
\end{equation}
or, equivalently,
\begin{equation}
L \frac{\partial}{\partial q'_{1}} \left( \frac{{\rm d} t}{{\rm d} t'} \right) + \frac{{\rm d} t}{{\rm d} t'} \frac{\partial L}{\partial q'_{1}} = 0. \label{cond2}
\end{equation}
We calculate the partial derivatives appearing on the left-hand side of the last equation separately. First, we have
\begin{equation}
\frac{\partial}{\partial q'_{1}} \left( \frac{{\rm d} t}{{\rm d} t'} \right) = \frac{\partial^{2} t}{\partial q'_{1} \partial t'} + \sum_{i = 1}^{n} \frac{\partial^{2} t}{\partial q'_{1} \partial q'_{i}} \dot{q}'_{i} = \frac{{\rm d}}{{\rm d} t'} \frac{\partial t}{\partial q'_{1}} = \frac{{\rm d} \xi}{{\rm d} t'}, \label{pt}
\end{equation}
where we have defined
\begin{equation}
\xi := \frac{\partial t}{\partial q'_{1}}. \label{xi}
\end{equation}

On the other hand, by virtue of the chain rule,
\begin{equation}
\frac{\partial L}{\partial q'_{1}} = \sum_{i = 1}^{n} \left( \frac{\partial L}{\partial q_{i}} \frac{\partial q_{i}}{\partial q'_{1}} + \frac{\partial L}{\partial \dot{q}_{i}} \frac{\partial \dot{q}_{i}}{\partial q'_{1}} \right) + \frac{\partial L}{\partial t} \frac{\partial t}{\partial q'_{1}}. \label{chr}
\end{equation}
The partial derivative of $\dot{q}_{i}$ with respect to $q'_{1}$ appearing in the last equation can be obtained as follows
\begin{eqnarray*}
\frac{\partial \dot{q}_{i}}{\partial q'_{1}} & = & \frac{\partial}{\partial q'_{1}} \lim_{\Delta t \rightarrow 0} \frac{\Delta q_{i}}{\Delta t} = \lim_{\Delta t \rightarrow 0} \frac{\partial}{\partial q'_{1}} \frac{\Delta q_{i}}{\Delta t} \\
& = & \lim_{\Delta t \rightarrow 0} \frac{\displaystyle \Delta t \frac{\partial \Delta q_{i}}{\partial q'_{1}} - \Delta q_{i} \frac{\partial \Delta t}{\partial q'_{1}}}{(\Delta t)^{2}} \\
& = & \lim_{\Delta t \rightarrow 0} \frac{\displaystyle \Delta t \; \Delta \frac{\partial q_{i}}{\partial q'_{1}} - \Delta q_{i} \; \Delta \frac{\partial t}{\partial q'_{1}}}{(\Delta t)^{2}} \\
& = & \frac{{\rm d}}{{\rm d} t} \frac{\partial q_{i}}{\partial q'_{1}} - \dot{q}_{i} \frac{{\rm d}}{{\rm d} t} \frac{\partial t}{\partial q'_{1}} \\
& = & \frac{{\rm d} \eta_{i}}{{\rm d} t} - \dot{q}_{i} \frac{{\rm d} \xi}{{\rm d} t},
\end{eqnarray*}
where we have made use of the definitions (\ref{xi}) and
\begin{equation}
\eta_{i} :=  \frac{\partial q_{i}}{\partial q_{1}'}. \label{etas}
\end{equation}
Hence, equation (\ref{chr}) takes the form
\begin{equation}
\frac{\partial L}{\partial q'_{1}} = \sum_{i = 1}^{n} \left[ \frac{\partial L}{\partial q_{i}} \eta_{i} + \frac{\partial L}{\partial \dot{q}_{i}} \left( \frac{{\rm d} \eta_{i}}{{\rm d} t} - \dot{q}_{i} \frac{{\rm d} \xi}{{\rm d} t} \right) \right] + \frac{\partial L}{\partial t} \xi. \label{chru}
\end{equation}
Substituting equations (\ref{pt}) and (\ref{chru}) into (\ref{cond2}) we obtain
\begin{equation}
L \frac{{\rm d} \xi}{{\rm d} t'} + \frac{{\rm d} t}{{\rm d} t'} \left\{ \sum_{i = 1}^{n} \left[ \frac{\partial L}{\partial q_{i}} \eta_{i} + \frac{\partial L}{\partial \dot{q}_{i}} \left( \frac{{\rm d} \eta_{i}}{{\rm d} t} - \dot{q}_{i} \frac{{\rm d} \xi}{{\rm d} t} \right) \right] + \frac{\partial L}{\partial t} \xi \right\} = 0, \label{casi}
\end{equation}
then, making use of the fact that
\[
\frac{{\rm d} \xi}{{\rm d} t'} = \frac{{\rm d} \xi}{{\rm d} t} \frac{{\rm d} t}{{\rm d} t'},
\]
and cancelling the common factor ${\rm d} t/{\rm d} t'$, we see that (\ref{casi}) is equivalent to equation (\ref{invinfs}).

Conversely, given a variational symmetry of the Lagrangian $L$ (represented by the functions $\xi$ and $\eta_{i}$), we can solve equations (\ref{xi}) and (\ref{etas}) and find a coordinate transformation $q_{i} = q_{i}(q'_{t}, t')$, $t = t(q'_{j}, t')$ such that $q'_{1}$ is an ignorable coordinate for the new Lagrangian $L'$. This is always possible as a consequence of the so-called straightening-out lemma (see, e.g., \cite{CP}). In Section 4 we give some explicit examples.

Now we can show that the momentum conjugate to the ignorable coordinate $q'_{1}$, which is conserved, coincides with the expression (\ref{conss}). We start from the expression (see equation (\ref{trla}))
\begin{equation}
p'_{1} = \frac{\partial L'}{\partial \dot{q}'_{1}} = \frac{\partial}{\partial \dot{q}'_{1}} \left( L \frac{{\rm d} t}{{\rm d} t'} \right) = L \frac{\partial}{\partial \dot{q}'_{1}} \left( \frac{{\rm d} t}{{\rm d} t'} \right) + \frac{{\rm d} t}{{\rm d} t'} \frac{\partial L}{\partial \dot{q}'_{1}}. \label{consmom}
\end{equation}
The first partial derivative contained in the right-hand side of the last equation is given by
\begin{equation}
\frac{\partial}{\partial \dot{q}'_{1}} \left( \frac{{\rm d} t}{{\rm d} t'} \right) = \frac{\partial}{\partial \dot{q}'_{1}} \left( \frac{\partial t}{\partial t'} + \sum_{i = 1}^{n} \frac{\partial t}{\partial q'_{i}} \dot{q}'_{i} \right) = \frac{\partial t}{\partial q'_{1}} = \xi \label{prov}
\end{equation}
and, in order to calculate the second one, making use of the chain rule, we have
\begin{equation}
\frac{\partial L}{\partial \dot{q}'_{1}} = \sum_{i = 1}^{n} \frac{\partial L}{\partial \dot{q}_{i}} \frac{\partial \dot{q}_{i}}{\partial \dot{q}'_{1}}. \label{cra}
\end{equation}
From the expression
\[
\dot{q}_{i} = \frac{{\rm d} t'}{{\rm d} t} \frac{{\rm d} q_{i}}{{\rm d} t'} = \frac{{\rm d} t'}{{\rm d} t} \left( \frac{\partial q_{i}}{\partial t'} + \sum_{j = 1}^{n} \frac{\partial q_{i}}{\partial q'_{j}} \dot{q}'_{j} \right),
\]
with the aid of equations (\ref{etas}) and (\ref{prov}), we obtain
\begin{eqnarray*}
\frac{\partial \dot{q}_{i}}{\partial \dot{q}'_{1}} & = & \frac{{\rm d} t'}{{\rm d} t} \frac{\partial q_{i}}{\partial \dot{q}'_{1}} + \frac{{\rm d} q_{i}}{{\rm d} t'} \frac{\partial}{\partial \dot{q}'_{1}} \left( \frac{{\rm d} t'}{{\rm d} t} \right) \\
& = & \frac{{\rm d} t'}{{\rm d} t} \eta_{i} + \frac{{\rm d} q_{i}}{{\rm d} t'} \frac{\partial}{\partial \dot{q}'_{1}} \left( \frac{{\rm d} t}{{\rm d} t'} \right)^{-1} \\
& = & \frac{{\rm d} t'}{{\rm d} t} \eta_{i} - \frac{{\rm d} q_{i}}{{\rm d} t'} \left( \frac{{\rm d} t}{{\rm d} t'} \right)^{-2}
\frac{\partial}{\partial \dot{q}'_{1}} \left( \frac{{\rm d} t}{{\rm d} t'} \right) \\
& = & \frac{{\rm d} t'}{{\rm d} t} \eta_{i} - \frac{{\rm d} q_{i}}{{\rm d} t'} \left( \frac{{\rm d} t'}{{\rm d} t} \right)^{2} \xi \\
& = & \frac{{\rm d} t'}{{\rm d} t} (\eta_{i} - \xi \dot{q}_{i}).
\end{eqnarray*}
Substituting this result into (\ref{cra}) we have
\[
\frac{\partial L}{\partial \dot{q}'_{1}} = \frac{{\rm d} t'}{{\rm d} t} \sum_{i = 1}^{n} \frac{\partial L}{\partial \dot{q}_{i}} (\eta_{i} - \xi \dot{q}_{i})
\]
and, therefore, the conserved momentum (\ref{consmom}) is given by
\[
p'_{1} = L \xi + \sum_{i = 1}^{n} \frac{\partial L}{\partial \dot{q}_{i}} (\eta_{i} - \xi \dot{q}_{i}),
\]
which agrees with equation (\ref{conss}).

\section{Noether--Bessel-Hagen symmetries}
If one is interested in finding conserved quantities, the condition $\partial L/\partial q_{1} = 0$ is unnecessarily restrictive. If
\begin{equation}
\frac{\partial L}{\partial q_{1}} = \frac{{\rm d} G}{{\rm d} t}, \label{gigno}
\end{equation}
for some function $G(q_{i}, t)$, then, from the Euler--Lagrange equations (\ref{lageq}), it follows that $p_{1} - G$ is conserved:
\[
\frac{{\rm d} (p_{1} - G)}{{\rm d} t} = 0.
\]

Certainly, it is easy to recognise the absence of a coordinate in a given Lagrangian, and, at first sight, it may seem difficult to discover a coordinate such that equation (\ref{gigno}) holds. However, in some cases, the physical or geometrical nature of the system helps to discover such a coordinate. For instance, in the case of a particle in a {\em uniform}\/ gravitational field, the standard Lagrangian is given (in Cartesian coordinates) by
\begin{equation}
L = \frac{m}{2} (\dot{x}^{2} + \dot{y}^{2} + \dot{z}^{2}) - mgz. \label{3grav}
\end{equation}
Even though $z$ is not ignorable, the assumed uniformity of the gravitational field suggests the existence of a conserved quantity associated with this coordinate. Indeed, we have
\[
\frac{\partial L}{\partial z} = -mg = - \frac{{\rm d} (mgt)}{{\rm d} t},
\]
which is of the form (\ref{gigno}), with $G = -mgt$ (and, therefore, $p_{z} + mgt$ is a constant of motion).

This generalisation of the concept of ignorable coordinate corresponds to the so-called divergence symmetries \cite{PO}, or Noether--Bessel-Hagen symmetries \cite{KS}, which are defined by functions $\eta_{i}(q_{j}, t)$, $\xi(q_{j}, t)$ such that
\begin{equation}
\sum_{i = 1}^{n} \left[ \frac{\partial L}{\partial q_{i}} \eta_{i} + \frac{\partial L}{\partial \dot{q}_{i}} \left( \frac{{{\rm d}} \eta_{i}}{{{\rm d}} t} - \dot{q}_{i} \frac{{{\rm d}} \xi}{{{\rm d}} t} \right) \right] + \frac{\partial L}{\partial t} \xi + L \, \frac{{{\rm d}} \xi}{{{\rm d}} t} = \frac{{{\rm d}} G}{{{\rm d}} t}, \label{nbh}
\end{equation}
for some function $G(q_{i},t)$. If $\eta_{i}(q_{j}, t)$, $\xi(q_{j}, t)$ satisfy equation (\ref{nbh}), then
\begin{equation}
\varphi(q_{i}, \dot{q}_{i}, t) = \sum_{i = 1}^{n} \frac{\partial L}{\partial \dot{q}_{i}} \eta_{i} + \xi \left( L - \sum_{i = 1}^{n} \frac{\partial L}{\partial \dot{q_{i}}} \dot{q}_{i} \right) - G \label{consbh}
\end{equation}
is conserved.

Following the steps given in the preceding section, one finds that equation (\ref{nbh}) is equivalent to the existence of a system of coordinates $(q'_{i}, t')$ in the extended configuration space such that
\begin{equation}
\frac{\partial L'}{\partial q'_{1}} = \frac{{\rm d} G}{{\rm d} t'}, \label{ext}
\end{equation}
where $G$ is the function appearing in the right-hand side of (\ref{nbh}).

The function $G$ can always be expressed in the form
\begin{equation}
G = - \frac{\partial F}{\partial q'_{1}}, \label{efe}
\end{equation}
where $F$ is a function of $(q'_{i}, t')$. In fact, $F$ is defined up to an arbitrary additive function of the $n$ variables $(q'_{2}, q'_{3}, \ldots, q'_{n}, t')$. Substituting (\ref{efe}) into (\ref{ext}) we find that
\begin{equation}
\frac{\partial}{\partial q'_{1}} \left( L' + \frac{{\rm d} F}{{\rm d} t'} \right)= 0. \label{ignmod}
\end{equation}

On the other hand, as is well known, two Lagrangians, $L$ and $\tilde{L}$, yield the same Euler--Lagrange equations, i.e.,
\begin{equation}
\frac{{\rm d}}{{\rm d} t} \frac{\partial L}{\partial \dot{q}_{i}} - \frac{\partial L}{\partial q_{i}} = \frac{{\rm d}}{{\rm d} t} \frac{\partial \tilde{L}}{\partial \dot{q}_{i}} - \frac{\partial \tilde{L}}{\partial q_{i}}, \label{lagequiv}
\end{equation}
for $i = 1, 2, \ldots, n$, if and only if there exists a function $F(q_{i}, t)$ such that $\tilde{L} = L + {\rm d} F/{\rm d} t$. Thus, equation (\ref{ignmod}) tells us that $q'_{1}$ is an ignorable coordinate for the Lagrangian $\tilde{L}' = L' + {\rm d} F/{\rm d} t'$, which is equivalent to $L'$ (in the sense of equation (\ref{lagequiv})).

Summarising, we have shown that if the functions $\xi$, $\eta_{i}$ satisfy equation (\ref{nbh}), for some function $G$, then there exist coordinates $(q'_{i}, t')$ such that $q'_{1}$ is an ignorable coordinate of the Lagrangian $L'$, or of a Lagrangian $\tilde{L'}$, equivalent to $L'$.

\section{Examples}
All the variational symmetries of a given Lagrangian can be obtained solving equation (\ref{nbh}), making use of the fact that $\xi$ and $\eta_{i}$ are functions of $q_{i}$ and $t$ only, and that equation (\ref{nbh}) must hold for all values of $q_{i}$, $\dot{q}_{i}$, and $t$, without imposing the equations of motion. In this manner, one finds, for instance, that in the case of the Lagrangian
\begin{equation}
L = \frac{m}{2} \left( \dot{x}^{2} + \dot{y}^{2} \right) - mgy, \label{exlag}
\end{equation}
the solution of equation (\ref{nbh}) contains eight arbitrary constants \cite{GFTC-AndradeMiron2013,GFTC2014}
\begin{eqnarray}
\xi & = & c_{3} + c_{7}t + c_{8}t^{2}, \nonumber \\
\eta_{1} & = & c_{1} + c_{4}t + c_{6} \left( {\textstyle \frac{1}{2}} gt^{2} + y \right) + {\textstyle \frac{1}{2}} c_{7} x + c_{8} xt, \label{gens} \\
\eta_{2} & = & c_{2} + c_{5}t - c_{6}x + c_{7}\left( {\textstyle \frac{1}{2}} y - {\textstyle \frac{3}{4}} gt^{2} \right) + c_{8} \left( yt - {\textstyle \frac{1}{2}} gt^{3} \right), \nonumber
\end{eqnarray}
where $c_{1}, \ldots, c_{8}$ are arbitrary real constants, and the corresponding function $G$ is given by
\begin{eqnarray}
\nonumber G & = & -c_{2}mgt + c_{4}mx + c_{5}m (y - {\textstyle \frac{1}{2}} gt^{2}) + c_{6} mgxt + c_{7} m ( - {\textstyle \frac{3}{2}} gyt + {\textstyle \frac{1}{4}} g^{2} t^{3}) \\
& & \mbox{} + c_{8} m ( - {\textstyle \frac{3}{2}} gt^{2} y + {\textstyle \frac{1}{2}}( x^{2} + y^{2}) + {\textstyle \frac{1}{8}} g^{2} t^{4}). \label{ge}
\end{eqnarray}
In order to simplify the calculations, we shall consider some particular solutions contained in (\ref{gens})--(\ref{ge}).

\subsection{Case $\xi = 0$, $\eta_{1} = t$, $\eta_{2} = 0$}
In the case where $c_{4} = 1$ is the only constant in equations (\ref{gens})--(\ref{ge}) different from zero, $\xi = 0$, $\eta_{1} = t$, $\eta_{2} = 0$, and $G$ is given by $G = mx$. Letting $x := q_{1}$, $y := q_{2}$ and, similarly, $x' := q'_{1}$, $y' := q'_{2}$, from the definitions (\ref{xi}) and (\ref{etas}) we have
\begin{equation}
\frac{\partial t}{\partial x'} = 0, \qquad \frac{\partial x}{\partial x'} = t, \qquad \frac{\partial y}{\partial x'} = 0. \label{cte1}
\end{equation}
A coordinate transformation satisfying these conditions is given by
\[
x = x' t', \qquad y = y', \qquad t = t',
\]
and the Lagrangian in the primed coordinates is (see equation (\ref{trla}))
\[
L'(q'_{i}, \dot{q}'_{i}, t') = {\textstyle \frac{1}{2}} m [(x' + t' \dot{x}')^{2} + (\dot{y}')^{2}] - mgy'.
\]

On the other hand, $G = m x't'$, which can be expressed in the form (\ref{efe}) with $F = - {\textstyle \frac{1}{2}} m (x')^{2} t'$, hence
\[
\tilde{L}' = L' + \frac{{\rm d} F}{{\rm d} t'} = {\textstyle \frac{1}{2}} m[(t'\dot{x}')^{2} + (\dot{y}')^{2}] - mgy',
\]
so that, in effect, $x'$ is ignorable and the conserved momentum conjugate to $x'$ is
\[
\frac{\partial \tilde{L}'}{\partial \dot{x}'} = m (t')^{2} \dot{x}' = m(t \dot{x} - x).
\]

\subsection{Case $\xi = t$, $\eta_{1} = {\textstyle \frac{1}{2}} x$, $\eta_{2} = {\textstyle \frac{1}{2}} y - {\textstyle \frac{3}{4}}g t^{2}$}
In the case where $c_{7} = 1$ is the only constant appearing in equations (\ref{gens})--(\ref{ge}) different from zero, $\xi = t$, $\eta_{1} = {\textstyle \frac{1}{2}} x$, $\eta_{2} = {\textstyle \frac{1}{2}} y - {\textstyle \frac{3}{4}}g t^{2}$, the function $G$ is given by $G = m ( - \frac{3}{2}gty + \frac{1}{4}g^{2}t^{3})$, and the definitions (\ref{xi}) and (\ref{etas}) give the set of equations
\[
\frac{\partial t}{\partial x'} = t, \qquad \frac{\partial x}{\partial x'} = \frac{x}{2}, \qquad \frac{\partial y}{\partial x'} = \frac{y}{2} - \frac{3}{4} g t^{2}.
\]
One can readily verify that a coordinate transformation satisfying these conditions is given by
\begin{equation}
x = a {\rm e}^{x'/2}, \qquad y = y' {\rm e}^{x'/2} - {\textstyle \frac{1}{2}} {\rm e}^{2x'} g(t')^{2}, \qquad t = {\rm e}^{x'} t', \label{ct7}
\end{equation}
where $a$ is a constant with dimensions of length.

In terms of the new coordinates, the function $G$ is expressed as
\[
G = - {\textstyle \frac{3}{2}} mg {\rm e}^{3x'/2} t' y' + mg^{2} {\rm e}^{3x'} (t')^{3}.
\]
This function can be written in the form (\ref{efe}), with, e.g.,
\[
F = mg {\rm e}^{3x'/2} t' y' - {\textstyle \frac{1}{3}} m g^{2} {\rm e}^{3x'} (t')^{3}.
\]
Then, according to equation (\ref{trla}),
\begin{equation}
\tilde{L}' = L' + \frac{{\rm d} F}{{\rm d} t'} = \frac{m}{2(1 + t'\dot{x}')} \left[ \left( {\textstyle \frac{1}{2}} a \dot{x}' \right)^{2} + \left( \dot{y}' + {\textstyle \frac{1}{2}}y' \dot{x}' \right)^{2} \right] \label{lag7}
\end{equation}
and we can see that $x'$ is indeed an ignorable coordinate.

The absence of the coordinate $x'$ in the Lagrangian (\ref{lag7}) means that this Lagrangian is invariant under the {\em finite}\/ translations $x' \mapsto x' + s$, where $s$ is an arbitrary real number. Making use of equations (\ref{ct7}) we find that, under these transformations, the original coordinate $x$ transforms according to $x \mapsto a {\rm e}^{(x' + s)/2} = x {\rm e}^{s/2}$. In a similar manner we obtain
\[
y \mapsto {\rm e}^{s/2} y + {\textstyle \frac{1}{2}r} gt^{2} ({\rm e}^{s/2} - {\rm e}^{2s}), \qquad t \mapsto t {\rm e}^{s}.
\]
Thus, as a by-product of the foregoing calculations, we obtain the explicit expression of a one-parameter group of transformations that leave invariant the original Lagrangian, $L$, up to the total derivative of a function $F(q_{i}, t)$. A similar identification can be obtained in the other examples of this section.

\subsection{Case $\xi = 1$, $\eta_{1} = 0$, $\eta_{2} = 0$}
In the case where the only constant appearing in equations (\ref{gens})--(\ref{ge}) different from zero is $c_{3} = 1$, we have $\xi = 1$, $\eta_{1} = 0$, $\eta_{2} = 0$, and $G = 0$. This symmetry corresponds to the absence of the variable $t$ in the original Lagrangian (\ref{exlag}); we consider this example in order to show that also this symmetry can be related to an ignorable coordinate $x'$.

A simple coordinate transformation satisfying equations (\ref{xi}) and (\ref{etas}) is
\[
t = x', \qquad x = t', \qquad y = y'.
\]
Then, a straightforward computation leads to the expression
\[
L' = \frac{m}{2} \left[ \frac{1}{\dot{x}'} + \frac{(\dot{y}')^{2}}{\dot{x}'} \right] - mgy' \dot{x}',
\]
that does not contain $x'$.

The conserved momentum conjugate to $x'$ is
\[
\frac{\partial L'}{\partial \dot{x}'} = - \frac{m}{2} \left[ \frac{1}{(\dot{x}')^{2}} + \frac{(\dot{y}')^{2}}{(\dot{x}')^{2}} \right] - mgy' = - \frac{m}{2} (\dot{x}^{2} + \dot{y}^{2}) - mgy,
\]
which can be recognised as minus the total energy (as one would expect, taking into account the meaning of the symmetry).

\section{Several ignorable coordinates}
Some Lagrangians possess more than one ignorable coordinate (for instance, the coordinates $x$ and $y$ are ignorable in the Lagrangian (\ref{3grav}) considered above) and one may ask, for instance, if the existence of two variational symmetries implies the existence of a coordinate system in the extended configuration space such that two of the coordinates are ignorable. The answer, in general, is no.

As an example, we may try to combine the variational symmetries considered in Sections 4.1 and 4.3; that is, we want to associate the symmetry corresponding to $\xi = 0$, $\eta_{1} = t$, $\eta_{2} = 0$, with an ignorable coordinate $x'$ (just as we did in Section 4.1), and, at the same time, the symmetry corresponding to $\xi = 1$, $\eta_{1} = 0$, $\eta_{2} = 0$, with an ignorable coordinate $y'$. This means that we look for a coordinate system $(x', y', t')$ such that (cf.\ equation (\ref{cte1}))
\begin{equation}
\frac{\partial t}{\partial x'} = 0, \qquad \frac{\partial x}{\partial x'} = t, \qquad \frac{\partial y}{\partial x'} = 0, \label{5.1}
\end{equation}
and, in analogous manner,
\begin{equation}
\frac{\partial t}{\partial y'} = 1, \qquad \frac{\partial x}{\partial y'} = 0, \qquad \frac{\partial y}{\partial y'} = 0. \label{5.2}
\end{equation}
However, these equations are not compatible. In fact, from equations (\ref{5.1}) and (\ref{5.2}) we have
\[
\frac{\partial}{\partial y'} \frac{\partial x}{\partial x'} = \frac{\partial t}{\partial y'} = 1,
\]
and
\[
\frac{\partial}{\partial x'} \frac{\partial x}{\partial y'} = \frac{\partial}{\partial x'} 0 = 0.
\]
Thus, it is impossible to find a coordinate system such that the chosen variational symmetries correspond to two ignorable coordinates. (Something similar happens in the case of a particle in a central field of force; in spite of the fact that the standard Lagrangian is invariant under the rotations about the three Cartesian axes, even in spherical coordinates, only the rotations about one axis can be associated with an ignorable coordinate.)

\section*{Acknowledgements}
One of the authors (I R-G) thanks PRODEP-SEP for financial support through a postdoctoral scholarship DSA/103.5/16/5800 and also the Sistema Nacional de Investigadores (M\'exico).


\begin{thebibliography}{9}
\bibitem{HS} Stephani H 1990 {\it Differential Equations: Their Solution Using Symmetries} (Cambridge: Cambridge University Press)
\bibitem{PO} Olver P J 2000 {\it Applications of Lie Groups to Differential Equations} 2nd edn (New York: Springer)
\bibitem{PH} Hydon P E 2000 {\it Symmetry Methods for Differential Equations: A Beginner's Guide} (Cambridge: Cambridge University Press)
\bibitem{GFTC-AndradeMiron2013} Torres del Castillo G F, Andrade Mir\'on C and Bravo Rojas R I 2013 Variational symmetries of Lagrangians {\it Rev. Mex. F\'is. E} {\bf 59} 140-7.
\bibitem{KS} Kosmann-Schwarzbach Y 2011 {\it The Noether Theorems: Invariance and Conservation Laws in the Twentieth Century} (New York: Springer) chapter 4
\bibitem{CP} Crampin M and Pirani F A E 1986 {\it Applicable Differential Geometry} (Cambridge: Cambridge University Press) chapter 6
\bibitem{GFTC2014} Torres del Castillo G F 2014 Point symmetries of the Euler-Lagrange equations {\it Rev. Mex. F\'is.} {\bf 60} 129-35.
\end{thebibliography}
\end{document}